# Strictly all-fiber picosecond Ytterbium fiber laser utilizing chirped-fiber-Bragg-gratings for dispersion control

Ori Katz, Yoav Sintov

*Nonlinear Optics Group, Electro-Optics Division, Soreq NRC, Yavne 81000, Israel*
*ori.katz@weizmann.ac.il, sintovy@soreq.gov.il*

**Abstract:** A compact, strictly all-fiber, picosecond pulse source based on ytterbium (Yb) doped fiber is described. Stable solitary mode-locking is obtained in a fiber-oscillator utilizing a carefully designed chirped fiber-Bragg-grating (C-FBG) for both dispersion control and spectral filtering. Self-starting is assured through the use of a fiber-coupled semiconductor-saturable-absorber-mirror (SESAM). The oscillator's 50MHz 3.8ps pulse-train output at 1064nm wavelength is amplified to 1.2W average power by an Yb doped fiber-amplifier, yielding 6.45ps parabolic pulses. Numerical simulations of the fiber oscillator design based on the modified nonlinear Schrödinger equation (NLSE), agree well with the experimental results.

Keywords: Mode-locked lasers; Optical fiber lasers; Semiconductor devices; Ytterbium;

**1. Introduction**

In recent years there has been an extensive growth in research activities focused on the development of ultrashort (<10ps) pulse sources based on rare-earth-doped-fibers [1]. This trend is motivated by the advantages of the short durations and high peak powers of ultrashort pulses for numerous practical applications. Examples for such applications are ultrafast spectroscopy, nonlinear microscopy, white-light continuum generation, laser micro-machining, coherent-control and nonlinear frequency conversions [1]. In addition, the inherent benefits of fiber lasers when compared to their solid-state lasers counterparts, are beneficial for such applications. These include smaller physical dimensions, reduced thermal management, increased stability to environmental conditions and relative ease of engineering. In particular, ytterbium ($Yb^{3+}$) doped fibers form an excellent gain medium for the generation and amplification of ultrashort optical pulses in the 1 micron wavelength range, due to their broad emission bandwidth, large saturation fluence and high optical to optical conversion efficiency.

However, in the development of short-pulse fiber sources in the 1μm wavelength range, a difficulty arises from the high value of normal material group-velocity-dispersion (GVD) of silica at wavelengths below 1.1μm. Without proper dispersion management, the mode-locking process in typical fiber-cavity lengths may become unstable and difficult to self-start, due to the combined effects of the normal GVD and the nonlinear Kerr effect [2, 3]. This fact dictates the introduction of anomalous GVD to the cavity in the form of dispersion compensating elements. The stabilization of the ultrashort pulse is then carried out through soliton-like pulse shaping mechanisms [2-4], or through self-similar parabolic pulse evolution [5]. Dispersion compensation is usually achieved by using bulk optical components such as prisms or diffraction gratings [6-8]. These solutions, however, have the major disadvantage of loosing the above mentioned "all-fiber" benefits of the laser system. Another relatively novel solution for dispersion management is the use of anomalous dispersion photonic band-gap (PBG) fibers [9,10]. This solution has the advantage of maintaining the all-fiber qualities, but results in increased cavity lengths when high values of anomalous dispersion are required. This may not pose a problem when femtosecond pulses are concerned (i.e. a small value of anomalous dispersion is needed) [10], but might limit the obtainable pulse repetition rates when picosecond-scale pulse durations are desired. Another drawback of the use of PBG fibers is the non-trivial splicing procedures of PBG fibers compared to standard silica fibers.

In contrast to the above mentioned dispersion-compensation schemes, the use of a C-FBG [11,12] is an exceptionally robust dispersion control scheme that maintains the all-fiber benefits, without compromising on any of the cavity characteristics. The C-FBG is a compact, reliable silica-fiber based component that can be engineered to meet a wide range of desired design parameters. The value of anomalous dispersion that is added to the fiber laser cavity is directly determined by the C-FBG chirp. Moreover, the controlled C-FBG reflection characteristics can be exploited for intracavity spectral filtering of the mode-locked pulses.

In this paper we present the experimental implementation of a compact, strictly all-fiber, picosecond pulse source. The laser is based on $Yb^{3+}$ doped fiber and utilizes an anomalous dispersion C-FBG for dispersion control and spectral-filtering. The mode-locking process is self-initiated and sustained through the use of a fiber-coupled semiconductor saturable absorber mirror (SESAM) [13]. The C-FBG dispersion and spectral reflection profile were carefully selected to allow solitary picosecond pulse formation, while minimizing, through spectral filtering, the energy of the spurious spectral soliton-sidebands [17], which are characteristic of dispersion-managed picosecond fiber-lasers [3, 12].

The combined use of a fiber-coupled SESAM as a mode-locker and a C-FBG for dispersion control, allows for a compact, environmentally stable all-fiber integrated system. The experimental results are backed by detailed numerical modeling of the laser oscillator design.

## 2. System design

The system design is illustrated in Fig. 1. The system is based on a passively mode-locked solitary all-fiber laser oscillator, which is amplified in a master oscillator power amplifier (MOPA) configuration by a two-stage Yb-doped fiber amplifier.

The oscillator is based on a previous published work [12], and its parameters are summarized in Table 1. The design is a linear cavity composed of single-mode fibers. The back mirror is a fiber-coupled SESAM, and the front mirror is a high reflectivity (~97%) anomalous dispersion C-FBG. The SESAM consists of an InGaAs quantum well absorber grown on top of a GaAs/AlGaAs Bragg-mirror on a GaAs substrate [13]. The SESAM is glued to a FC/PC connector at the end of the fiber cavity. The C-FBG is a linearly-chirped Bragg-reflector with 1.56nm FWHM reflection bandwidth (Fig. 2). The C-FBG is fabricated in a photosensitive fiber, through UV exposure of an e-beam generated phase mask [14]. An all-fiber polarization beam-combiner/splitter (PBS) is used as an output-coupler, and the output coupling is controlled via a miniature inline fiber polarization-controller (PC). The gain is provided by a 50cm long $Yb^{3+}$ doped single-mode fiber, pumped by a 976nm fiber-coupled diode through a 976/1064nm WDM-coupler. The total fiber length of the laser cavity is 2m, yielding a pulse repetition rate of 50MHz.

The oscillator output is connected to a two stage non-polarization maintaining fiber amplifier. The first amplifier stage is based on a 3.5m long, single-mode, double-clad Yb doped fiber, with a doping level of $3.4 \cdot 10^{25}$ m$^{-3}$. The second amplifier stage is based on a 4m long, large-mode-area (LMA), multi-mode double-clad Yb doped fiber supporting a beam quality of $M^2$=1.5, with doping levels of $1.4 \cdot 10^{25}$ m$^{-3}$. The two amplification stages are connected through a fiber isolator. The amplifier is cladding-pumped by two 976nm fiber-coupled diodes, supplying 1.25W for the first amplification stage and 2.5W for the second stage. The pump power is coupled into the inner clad of the double-clad fiber at each end of the amplifier, by a robust strictly all-fiber side coupling arrangement with a coupling efficiency of 90%. This novel side pumping technique is based on attaching the tapered end of the pump fiber to a stripped section of the amplifier's double-clad fiber [15]. In the attachment section, both pump and DC fibers are connected by fusing or gluing them together. The pump coupling occurs at the attachment section, which is the pump fiber's tapered section.

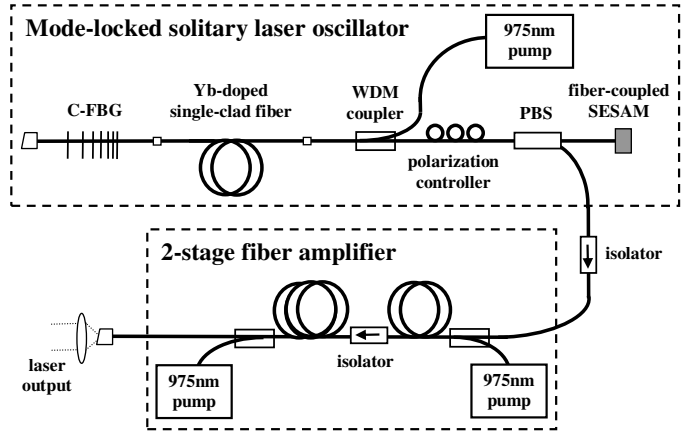

Fig.1. Picosecond pulse source MOPA configuration.

Table 1. Oscillator parameters

| PARAMETER | Value |
| --- | --- |
| Yb doped fiber Mode-field-diameter (MFD) | 5.4 μm |
| Yb doped fiber absorption (@976nm) | 219 dB/m |
| SESAM saturable / non-sat. absorption | 6% / 4% |
| SESAM saturation fluence | 60 μJ/cm$^2$ |
| SESAM relaxation time constant | ~500 fs |
| SESAM Bragg-mirror reflection bandwidth | 1040-1100nm (>90%) |
| C-FBG reflection | 96.7% |
| C-FBG dispersion (anomalous) | 8 ps/nm (-4.77ps$^2$) |
| C-FBG reflection bandwidth | 1.56 nm (FWHM) |

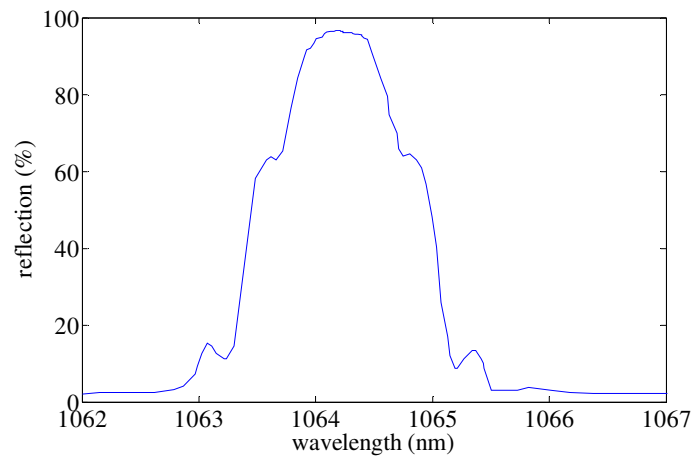

Fig.2. Measured C-FBG spectral reflection profile

## 3. Determining the C-FBG parameters

In order to determine the optimal C-FBG parameters for realizing an all-fiber picosecond pulse source, a numerical model of the fiber oscillator based on the pulse propagation equation was developed [4,12]. In the developed numerical simulation the electric-field is propagated numerous consecutive roundtrips inside the cavity, while experiencing the distributed effects of the fiber dispersion, Kerr nonlinearity, gain and loss, and the discrete effects of the SESAM saturable and non-saturable absorptions and C-FBG dispersion and spectral filtering. The field propagation is performed using the split-step-Fourier-method [4], where the nonlinear effects of SPM and saturable-absorption are evaluated in the time domain, and the linear effects (e.g. dispersion, spectral filtering) are evaluated in the frequency domain. Noise generation was not included in the numerical model.

The numerical simulation's results indicate that a soliton-like pulse formation process should initiate from slow (i.e. >100ps) intracavity power fluctuations whenever a sufficient amount of anomalous dispersion is introduced to the cavity by the C-FBG [12]. This is in agreement with the well-known theory of solitary mode-locking [2,3]. The steady state solitary-pulse duration is a function of the total anomalous dispersion in the cavity, which is determined by the C-FBG dispersion value. A larger value of total anomalous dispersion results in longer solitary pulses and vice-versa [2,12]. Numerous simulation runs with the presented design parameters indicated that a C-FBG dispersion value of about 8ps/nm is required to achieve solitary pulses of 4ps duration. At this parameter range if no spectral filtering is introduced by the C-FBG, strong narrow spectral sidebands are present at the oscillator output [3,12,17]. These spurious sidebands are characteristic of dispersion-managed fiber-lasers, and are the product of the dispersive field (continuum radiation) radiated by the perturbed solitary pulse along its propagation in the cavity [3,17]. The presence of these soliton-sidebands results in an unwanted CW-like noise at the mode-locked oscillator output. These spurious sidebands can be filtered-out in the cavity by a careful design of the C-FBG spectral reflection profile. The C-FBG reflection bandwidth should be wide enough to allow the formation of the solitary pulse, but as narrow as possible to minimize the energy of the spurious spectral soliton-sidebands [17]. The exact value for the C-FBG reflection bandwidth, taking into account the above considerations was determined using the numerical simulation, and is 1.5±0.1nm (FWHM). The experimentally fabricated C-FBG reflection profile has a reflection bandwidth of 1.6nm (Fig. 2).

## 4. Experimental results

### 4.1 Oscillator output characteristics

After initial adjustment of the PC position, stable solitary mode-locking self-starts at a threshold pump power of 14mW. At this power level the output pulses autocorrelation is 7ps wide, corresponding to a $sech^2$ pulse width of 4.5ps (Fig. 3). The average output power of the oscillator is approximately 1mW. Increasing the pump power results in pulse shortening. At a pump power of 22mW the pulses autocorrelation is 5.8ps wide, corresponding to 3.8ps $sech^2$ pulses, and the average output power is 2mW. At higher pump powers multiple pulsing and multisoliton formation [16] are observed. The pulses' autocorrelation traces at the above mentioned pump powers are plotted in Fig. 3. The autocorrelation traces match a sech profile, typical for soliton fiber lasers [3,4].

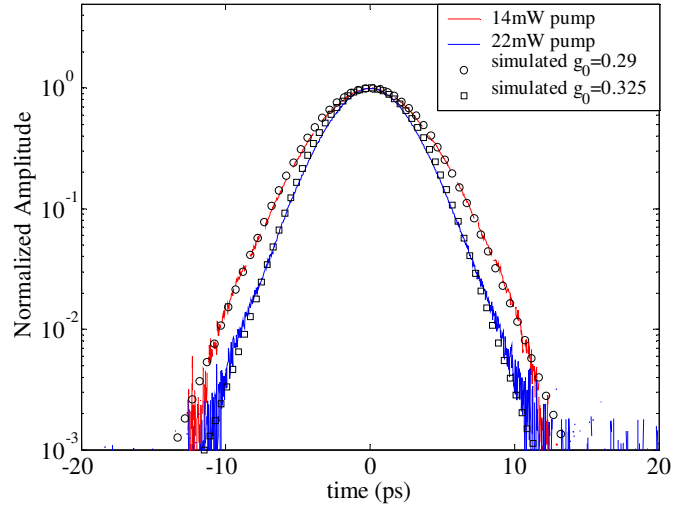

Fig. 3. Oscillator output autocorrelation traces: Solid/dashed curves - measured autocorrelation, open circles/squares - numerical simulation results. $g_0$ is the simulated small-signal amplitude roundtrip gain in the cavity, corresponding to the experimental pump-powers of 14mW and 22mW [12].

The spectral profile of the oscillator output at the above mentioned pump powers is plotted in Fig 4. At a pump power of 22mW the spectrum bandwidth is 0.4nm (FWHM) wide. This corresponds to a time bandwidth product of 0.4, which is close to the transform limited value of 0.315 of sech pulses, and indicates a small chirp of the output pulse. This small chirp is in agreement with our numerical simulation results, which suggest it to be a positive one. The chirp is positive, even though the net cavity dispersion is strongly anomalous, due to the fact that the output coupler is located at approximately half the round-trip length from the dispersion compensating element, where the pulses are nearly transform limited [2], and the pulses are measured after a ~1m long output delivery fiber having normal dispersion.

As apparent from Fig. 4, the characteristic soliton-sidebands [3, 12, 17] are strongly suppressed in the presented system due to the carefully designed C-FBG reflection profile, as discussed in the previous section. This fact accounts for the absence of temporal wings in the pulse autocorrelation trace (Fig. 3).

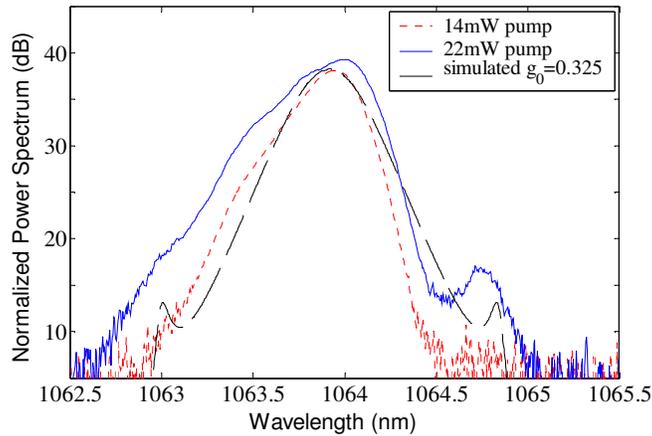

Fig. 4. Oscillator output spectral profiles under different pump powers. Solid/dotted curves – experimentally measured spectral profiles, dashed curve - numerical simulation result for the output spectral profile, under a small-signal amplitude roundtrip gain of $g_0$=0.325.

The numerical simulation results for the steady-state temporal and spectral pulse profiles relying on the laser parameters of Table 1, are in good agreement with the experimentally measured ones. To illustrate this, we have plotted two simulated autocorrelation traces on top of the measured ones in Fig. 3, and one spectral trace in Fig 4. The simulated round-trip amplitude gain parameter $g_0$ [12] was assumed to be 0.29 and 0.325 corresponding to the experimental pump powers of 14mW and 22mW, respectively. We attribute the difference between the simulated and measured output spectral profiles, to the limited accuracy in the modeling of the experimentally used C-FBG spectral reflection profile, which was numerically modeled as a nearly rectangular filter with smoothed edges of constant slope.

As previously mentioned, the oscillator output coupling is governed by the PC. Experimentally, the PC was adjusted to achieve self-starting mode-locking with maximum output power. The actual output-coupling value was estimated from the experimentally obtained output-powers using the numerical simulation, and is 0.25.

*4.2 MOPA output characteristics*

The MOPA output power characteristics measured at discrete pump power levels is shown in Fig 4. In all of the measurements the oscillator average output power was 2mW and the first stage pump power was 1.25W. An output power of 1.2W is achieved at a total pump power of 3.75W. The slope efficiency is 68% with respect to the second stage pump power.

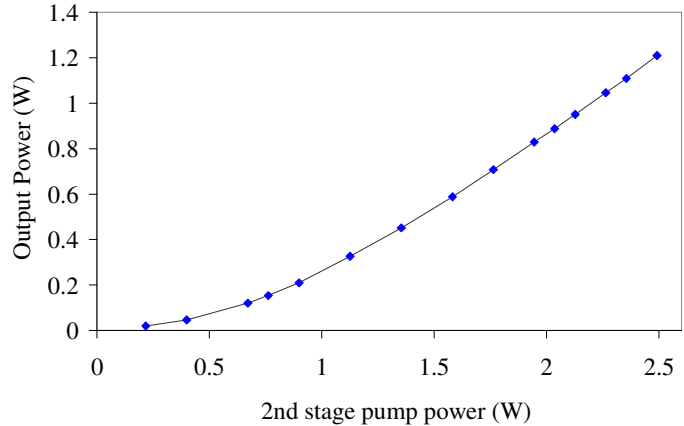

Fig. 5. MOPA output power characteristics. The oscillator output average power is 2mW and the pump power of the first amplification stage is 1.25W.

The spectral and autocorrelation profiles of the amplifier output are plotted in Fig. 6 and Fig. 7. The combined action of gain, nonlinearity and normal dispersion along the amplification stages should result in the generation of nearly linearly chirped parabolic pulses [18]. This was experimentally observed, as can be confirmed from the agreement of the solid measured autocorrelation trace and the dotted theoretical parabolic-pulse autocorrelation trace in Fig. 7. A similar result was also obtained from numerical simulation of the amplifier design. At 1.2W average output power, the amplified pulses have an autocorrelation width of 8ps, corresponding to a pulse duration of 6.45ps assuming a parabolic shape. The spectral width of the output at 1.2W average output power is 3nm (FWHM). The degree of polarization of the amplified output is between 80% to random polarization and is sensitive to the output fiber alignment. This results from the inherent birefringence variation in the non polarization-maintaining fiber stages of the amplifier.

The nearly linear chirp of the amplified pulses allows their recompression through an anomalous dispersion compensating element. This was done experimentally by using a grating pair with 1200 lines/mm and diffraction efficiency of 65%, placed at a separation distance of 9.2cm. The obtained de-chirped pulses possessed an autocorrelation width of 1.72ps (Fig. 7). Potentially higher compression efficiencies (i.e. lower losses) may be achieved by replacing the grating-pair compressor with an anomalous dispersion air-core PBG fiber, spliced to the amplifier output [19]. This will allow an all-fiber compressed pulse source, and will be studied in a future work.

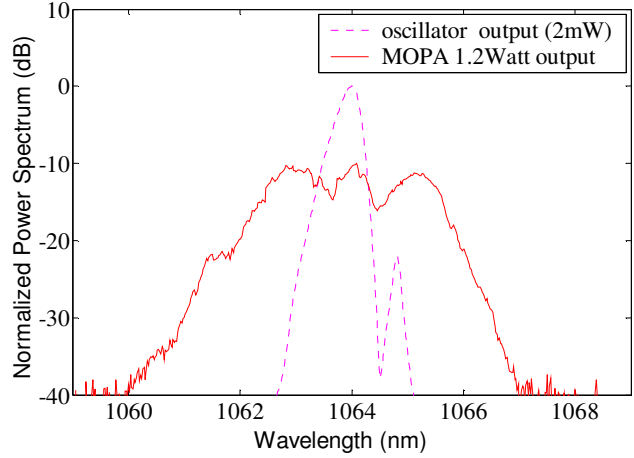

Fig 6. MOPA output spectral profiles.

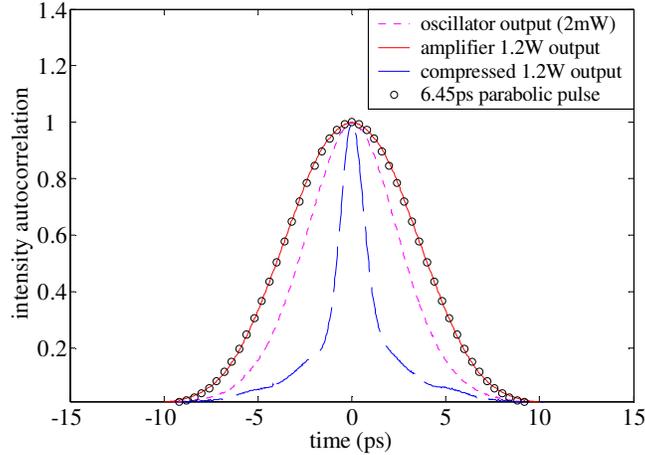

Fig. 7. MOPA output autocorrelation traces.

## 5. Conclusion

We have demonstrated a compact, strictly all fiber picosecond mode-locked laser, utilizing a C-FBG for spectral filtering and dispersion control. Stable self-starting mode-locked operation was accomplished by the use of a fiber-coupled SESAM as the cavity back mirror. The C-FBG was carefully designed to suppress the parasitic spectral soliton-sidebands typically observed in picosecond solitary mode-locked fiber-lasers. The spectral filtering feature is a unique quality of the C-FBG dispersion control method. The presented oscillator design is a promising scheme for a strictly "all-fiber" environmentally stable picosecond oscillator operating in the 1μm wavelength range. The polarized output of the laser oscillator was amplified in a non-polarization maintaining, all-fiber amplifier configuration producing 6.45ps nearly linearly-chirped parabolic pulses with an average power of 1.2W. The presented MOPA simple and robust design has promising features for numerous "all-fiber" 1μm range mode-locked practical applications, such as laser micro-machining and nonlinear frequency-conversions. The MOPA output pulses can be recompressed to about 1ps width using an anomalous-dispersion compressor.

Our numerical results show that various pulse widths in the picosecond regime (e.g. 1ps-100ps) can be obtained in the presented system design by changing the C-FBG dispersion value (0.5ps/nm-700ps/nm, respectively). In order to achieve sub-picosecond solitary pulses in the presented design, a C-FBG with <0.5ps/nm anomalous dispersion and a >5nm reflection bandwidth is required. Unfortunately, there is a fundamental difficulty in the fabrication of C-FBGs that combine such small dispersion values, large reflection bandwidths and high reflectivity values. According to our understanding, reflection values of more than 10% for such low-dispersion and high bandwidth are not feasible, and therefore require a different oscillator. However, a C-FBG with a low (<20%) value of reflection and a wider reflection bandwidth might be used in a similar experimental configuration as a high value output-coupler, achieving high energy pulses through self-similar parabolic-pulse evolution [7]. The self-similar parabolic pulse regime of operation does not require high reflectivity values as the solitary regime does, and will be studied in future work.

## Acknowledgement

The authors thank V-gen electro-optics for supplying the fiber-amplifier

## References and links